\documentclass[%
amsmath,amssymb,
aps,prl,twocolumn%
]{revtex4-2}

\usepackage{graphicx}
\usepackage{dcolumn}
\usepackage{bm}
\usepackage{xcolor}
\usepackage{braket}
\usepackage{ragged2e}
\usepackage{mwe}
\usepackage{tcolorbox}
\usepackage{bbold}
\usepackage{tkz-euclide}
\usepackage{wasysym}
\usepackage[T1]{fontenc}
\usepackage{blindtext, rotating}

\usepackage[unicode=true,colorlinks=true]{hyperref}
\hypersetup{linkcolor=blue,citecolor=blue,urlcolor=blue}

\newcommand{\bvec}[1]{{\bm{#1}}}

\newcommand{\expec}[1]{{\langle{#1}\rangle}}

\newcommand{\eu}{\mathrm{e}} 
\newcommand{\du}{\mathrm{d}} 

\begin{document}


\title{Kinetically induced order from mobility-constrained excitations}
\author{Aprem P. Joy}
\email{ajoy@uni-koeln.de}
\email{aprempjoy@gmail.com}

\author{Urban F.P. Seifert}
\affiliation{%
	Institute for Theoretical Physics, University of Cologne, Cologne, Germany
}%

\date{\today}

\begin{abstract}
A fascinating feature of strongly correlated systems is that excitations may carry nontrivial quantum numbers under emergent symmetries which, in some cases, may strongly constrain their motion.
We show that such constrained excitations can play a central role in driving symmetry-breaking long-range order through a kinetic mechanism: coupling to a local order-parameter field activates the mobility of otherwise constrained excitations, and condensation of the order parameter then enables a large gain in kinetic energy, thereby stabilizing a phase with long-range order.
We illustrate this mechanism in two settings:
(i) a dimerization instability in a spin chain with a magnetization dipole conservation law, and (ii) itinerant magnetism of electrons in the Kitaev-Kondo model, where charge carriers couple to local moments that form a spin liquid.
This mechanism becomes operative whenever a finite density of mobility-constrained excitations is present, either through non-equilibrium preparation or thermal activation. In the latter case, we argue that increasing temperature can, counterintuitively, promote order by enhancing the kinetic-energy gain of thermally activated excitations.
\end{abstract}

\maketitle

{\em Introduction.---}%
Spontaneous symmetry breaking and long-range order in many-body systems are typically driven by interactions. However, remarkably, in some systems, long-range order can have a kinetic origin.
A paradigmatic example is Nagaoka-Thouless ferromagnetism where the motion of a single hole in a Hubbard model with an infinite onsite repulsion favors a ferromagnetic background in order to maximize its kinetic energy~\cite{nagaoka66,thouless1965exchange}.
Generalizations of this mechanism demonstrate that kinetic effects can qualitatively alter conventional ordering tendencies~\cite{haertershastry,ciorciaro2023kinetic,dehollain2020nagaoka,shraimanprl1988,andersonrvbinterlayer}.

In this work, we show that similar phenomena can emerge in certain quantum many-body systems with mobility-constrained excitations.
These constraints can arise, for example, from higher-moment conservation laws or emergent gauge structures.
When the system is coupled to additional degrees of freedom which can undergo symmetry breaking,
the constraints responsible for quasiparticle immobility become relaxed, thereby allowing these excitations to gain kinetic energy.
Crucially, this gain in energy can outcompete the energetic cost of condensing an appropriate order parameter.

We explicitly demonstrate this in two example systems: a spin-1 chain with a magnetization-dipole conservation law, and the Kitaev honeycomb spin model.
While the former serves as a toy model for a fractonic system with non-ergodic behavior~\cite{pai19,sala20,feldmeier20,han24,adler24,lake23}, the latter is a paradigmatic example of an exactly solvable quantum spin liquid described by the deconfined phase of an emergent gauge theory.
First, we show that when the dipole-conserving spin chain is coupled to a lattice distortion/dimerization mode, formerly immobile single-spin excitations become dispersive quasiparticles once a finite dimerization develops, producing a Peierls-like instability.
Second, we study the Kitaev honeycomb model coupled to a band of itinerant electrons. When a finite density of point-like excitations of the emergent gauge field (``visons'') is present, it is energetically preferable for the electrons to develop a finite magnetization to maximize the kinetic energy of the visons.

\begin{figure}
    \centering
    \includegraphics[width=\columnwidth]{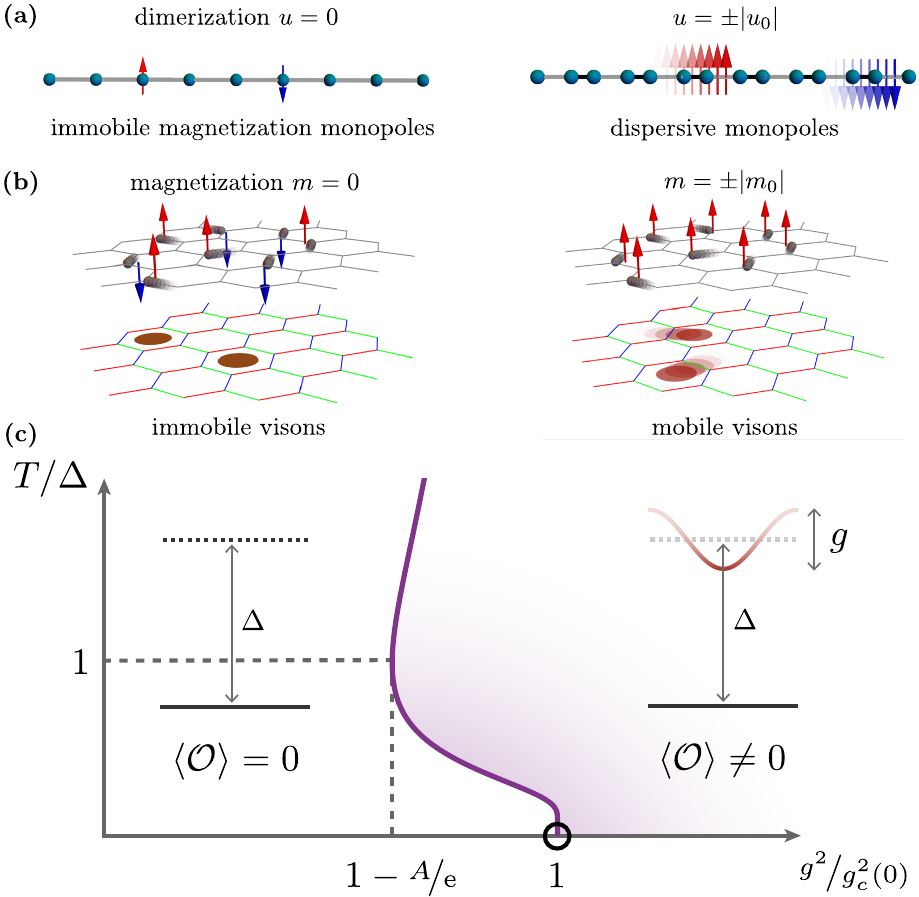}
    \caption{(a) In a spin chain with a magnetization dipole conservation law, a dimerization transition enables the mobility of single-particle excitations. (b) Vison excitations of the emergent gauge field in the Kitaev spin liquid acquire a dispersion when coupled to a magnetized conduction band.
   (c) Mean-field phase diagram: an order parameter $\langle \mathcal O\rangle$ activates the mobility of gapped excitations, yielding a kinetic energy gain $\propto g\langle\mathcal{O}\rangle$. Once this gain overcomes the cost of ordering, symmetry breaking occurs. In the dilute regime, $T \ll \Delta$, thermally activated excitations thus favor ordering, leading to a decreasing critical coupling $g_c$ with increasing temperature [see Eq.~\eqref{eq:relative-gc-T}].
    }
    \label{fig:introfig}
\end{figure}
Since a finite density of mobility-constrained quasiparticles can be thermally activated, we propose that the ordering-activated mobility can lead to an unusual scenario where \emph{increasing temperature} leads to a state with symmetry-breaking long-range order.
Such a temperature-enabled ordering is reminiscent of the celebrated Pomeranchuk effect~\cite{pomeranchuk1950theory,plazanet2004freezing} (see also Refs.~\onlinecite{hawashin25,han2025entropic}) and order-by-disorder phenomena~\cite{villain1980order}, both of which are entropic in origin.
More explicitly, our calculations in the dilute density limit (valid for temperature $T$ smaller than the excitation gap $\Delta$) show that the kinetic mechanism for ordering leads to a temperature-dependent {\em reduction} of the critical coupling of the form
\begin{equation} \label{eq:relative-gc-T}
    \frac{g^2(T) - g^2(0)}{g^2(0)} \approx - A \frac{\Delta}{T} \eu^{-\Delta/T},
\end{equation}
where $A$ is some numerical constant. A graphical summary of our results is presented in Fig.~\ref{fig:introfig}.


{\em Dipole conservation and dimerization instability.---}%
As a first example, we consider a chain of $N$ sites hosting $S=1$ moments with the Hamiltonian $H = H_J + H_D$~\cite{sala20}, where
\begin{align} \label{eq:h-dipole}
    H_D =& D \sum_i (S^z_i)^2\\ \nonumber 
   H_J =& -J \sum_i \left(  S^+_{i-1} (S^-_{i})^2 S^+_{i+1} + \mathrm{h.c.} \right).
\end{align}
For $D\gg J \geq  0$,  the system has a unique ground state $\ket{0} = \prod_{i} \ket{0_i}$ with energy $E_0 = 0$.
The lowest-lying excitations are (dressed) single-particle states $\ket{(\pm,i)} = \ket{\dots 0_{i-1} \pm_i 0_{i+1} \dots} + \mathcal{O}(J/D)$, with energy $E_{\mathsf{s}} = D + \mathcal{O}(J^2/D) > 0$.
Crucially, for \emph{any} finite $J$ (i.e.~arbitrarily high order in perturbation theory in $J/D \ll 1$), these states are fully localized: the Hamiltonian conserves not only the magnetization $M^z = \sum_i S^z_i$, but also the magnetization dipole moment $P^z = \sum_i (i-i_0) S^z_i $ (here, $i_0$ is an arbitrary reference site).
Thus, since $\ket{(\pm,i)}$ and $\ket{(\pm,i+1)}$ lie in different sectors of $P^z$, no mobility of single-particle states is allowed.

In the second-lowest-lying sector, corresponding to (dressed) two-particle excitations, most states (such as $S^+_i S^+_j \ket{0}$) are adiabatically connected to two single-particle excitations with energy $+2D + \mathcal{O}(J^2/D)$ which remain immobile by dipole symmetry.
A notable exception is given by the dipole states $\ket{(\leftarrow,i)} \equiv \ket{\dots 0_{i-1} +_{i} -_{i+1} 0_{i+2} \dots}$ and similarly $\ket{(\rightarrow,i)}$ \footnote{Note that dipole symmetry also allows for the mobility of extended dipoles such as $\ket{\dots 0_{i-2} +_{i-1} 0_i -_{i+1} 0_{i+2} \dots}$, but in the model at hand, they remain immobile owing to the locality of interactions in Eq.~\eqref{eq:h-dipole}.}, which are dispersive at any finite $J \neq 0$:
As shown explicitly in the Supplemental Material~\cite{suppl}, one may construct \emph{exact} momentum eigenstates of $H$ with energy $E_{\mathsf{d}} = 2D - 8J \cos k$, where $\mathsf{d} \in \{ \leftarrow,\rightarrow \}$ and $k \in (-\pi,\pi]$.

We now couple the system to an additional degree of freedom.
For concreteness, we consider a lattice distortion field $u_i$ which couples to the local moments via a magnetoelastic coupling of the form
\begin{equation} \label{eq:h_g}
    H_g[u_i] = g \sum_i u_i \vec{S}_i \cdot \vec{S}_{i+1}.
\end{equation}
We proceed by treating $u_i$ in a mean-field approximation (this is formally justified when the field is slow compared to spin dynamics, $\omega_0 \ll D,J$), focusing on the wavevector-$\pi$ component, $u_i \to u (-1)^i$.
Crucially, at finite $u \neq 0$, the coupling \eqref{eq:h_g} spoils the dipole conservation symmetry, $[H_g[u], P^z] \neq 0$, and in this case single-particle excitations are no longer constrained to be immobile.

The energy cost for developing a finite (mean) field configuration $u \neq 0$ can be modeled via the elastic energy $E_\mathrm{el.} = \rho N u^2/2$ (with an elasticity constant $\rho$), while the spin sector is expected to gain energy at finite $u$. Given this competition, is it energetically favorable to develop a finite dimerization?
The energy functional for the mean field $u$ is given by
\begin{equation} \label{eq:e-u}
    E[u]= E_\mathrm{el.}[u] + \braket{\psi[u]| H + H_g[u] |\psi[u]},
\end{equation}
The expectation value can be evaluated using perturbation theory in $gu \ll D, J$, where the leading non-zero contribution arises from mixing between the ground state and the band of dispersive dipole excitations, $\braket{\psi[u]| H + H_g[u] |\psi[u]} \approx - N \times (gu)^2/(D+4J)$.
Thus, for small $g$, the energy functional \eqref{eq:e-u} is minimized for $u = 0$ and no instability occurs.

This energetic competition changes drastically if we consider the system in an excited state,~i.e.~in the presence of one mobility-restricted single-particle excitation:
upon coupling to the distortion field $H_g[u]$, the $(2N)$-fold degeneracy of the states $\ket{(\mathsf{s},i)}$ (with $\mathsf{s} = \pm$) is lifted in first-order perturbation theory, giving rise to two \emph{dispersive} bands with $E_{\mathsf{s}}(k) = D - 2 |g u| \cos k$, $k \in (-\pi,\pi]$.
With one single-particle excitation occupying the band bottom with $k=0$, we arrive at the (perturbative) energy functional
\begin{equation} \label{eq:energy-linear}
    E[u] = \frac{N \rho}{2} u^2 - 2 |g| |u| + D + \mathcal{O}\left((gu)^2 /D \right).
\end{equation}
We conclude that there is a linear (non-analytic) gain in energy due to the dimerization activating the dispersion of the formerly immobile particle.
For any finite $g \neq 0$ and $N < \infty$, this outcompetes the elastic energy cost and the system spontaneously dimerizes with $u_\pm = \pm  \frac{2 |g|}{N \rho}$.
Note that accounting for the perturbative dressing of single-particle states amounts to replacing $g \to \tilde{g} = g(1 - 2 J/D + \mathcal{O}(J^2/D^2))$, and higher-order corrections in Eq.~\eqref{eq:energy-linear} only weakly renormalize the value $u_\pm$ provided that $g^2 \ll N \rho D$.

While the \emph{extensive} elastic energy contribution dominates over the kinetic energy gain of a \emph{single} excitation in the thermodynamic limit, the latter becomes important when a finite \emph{density} of excitations is present.
These could emerge via non-adiabatic state preparation protocols in cold-atom quantum simulator platforms~\cite{polkovnikov11}, or via thermal activation of excitations at finite temperature $T = \beta^{-1} >0$.
We demonstrate in the Supplemental Material~\cite{suppl} that, within the mean-field approximation, one may find an instability toward dimerization with a critical coupling of the form in Eq.~\eqref{eq:relative-gc-T}. However, long-range order in $1d$ will be destroyed by thermal fluctuations via proliferating domain walls at $T>0$~\cite{mostovoy96,weber16}.

{\em Fractionalized Fermi liquid in the Kitaev-Kondo model: vison excitations.---}%

We now turn to a two-dimensional system and consider the Kitaev-Kondo model~\cite{seifertvojta,choiRosch,colemanKondo,bunneyPRL,lundemo2024}, where itinerant electrons ($c_{i,\rho}$ with $\rho=\uparrow,\downarrow$) interact via an onsite Kondo coupling $J$ with spin-1/2 local moments $\vec{S}_i$ that form a Kitaev honeycomb model~(to connect with previous results on the Kitaev model, we henceforth find it convenient to directly work with the Pauli operators $\vec{\sigma}_i = 2 \vec{S}_i$) \cite{Kitaev06}.
The Hamiltonian is given by $H=H_t + H_K + H_J$ with
\begin{subequations}\begin{align}
    H_t &= \sum_{\langle ij\rangle,\rho=\uparrow,\downarrow} -t\left(c_{i\rho}^\dagger c_{j\rho}+\text{h.c.}\right)-\mu \sum_i c_{i\rho}^\dagger c_{i\rho} \label{eq:h-t} \\
    H_K &= -K\sum_{\langle ij\rangle_\alpha} \sigma_i^\alpha \sigma_j^\alpha \label{eq:h-k}\\ 
    H_{J} &= J\sum_i \vec{\sigma}_i \cdot \vec s_i. \label{eq:h-j}
\end{align}\end{subequations}
where $\vec{s}_i = \frac{1}{2} c_{i,\rho}^\dagger \vec \tau_{\rho,\rho'} c_{i,\rho'}$ is the spin density of itinerant electrons, and $\langle ij\rangle_\alpha$ denotes a nearest-neighbor bond of the honeycomb lattice connecting $\vec \sigma_i$ and $\vec \sigma_j$ which interact via the $\alpha=x,y,z$ component of spin.
Throughout this work, we will focus on a ferromagnetic Kitaev coupling, $K>0$.

For $J=0$, the electrons decouple from the exactly solvable Kitaev model, which hosts gapless Majorana fermions coupled to a $\mathbb{Z}_2$ gauge field whose conserved plaquette fluxes are $W_p =  \prod_{\langle ij \rangle \in \hexagon_p} \sigma^\alpha_i \sigma^\alpha_j=\pm 1$~\cite{Kitaev06,liu24}. The flux-free sector, $W_p=+1$, contains the ground state, while flipping a plaquette flux creates a localized excitation with energy $\Delta_v\approx0.15K$, dubbed ``vison''. Since all $W_p$ are conserved at $J=0$, these visons are immobile and thus analogous to the localized single-particle excitations of the dipole-conserving chain.
For small Kondo couplings $J \ll |K|$, the Kitaev spin liquid and its deconfined $\mathbb{Z}_2$ gauge structure are protected by the vison gap.
The model thus realizes a fractionalized Fermi liquid (FL$^\ast$) phase, featuring both electronic quasiparticles and fractionalized excitations~\cite{senthil03,senthil04,seifertvojta}.


{\em Vison mobility from spontaneous magnetic ordering.---}%
We now investigate magnetic ordering instabilities inside the FL$^\ast$ phase via a mean-field decoupling of the Kondo interaction.
With the mean-field parameters $\vec m=\expec{\vec s_i}$ and $\vec M=\expec{\vec \sigma_i}$, the Hamiltonian is written as
\begin{align}
H_{\text{mf}} =& H_t+J\sum_i \vec{M} \cdot \vec{s}_i+ H_K+J \sum_i \vec m\cdot \vec{\sigma}_i\\ \nonumber
&-NJ\vec m \cdot \vec M.
\label{eq:mean-field}
\end{align}
where $H_t$ denotes the electronic hopping, $H_K$ denotes the Kitaev term in Eq.~\eqref{eq:h-k}, and $N$ is the number of sites.
Thus, $H_{\text{mf}}$ decomposes into a free-fermion piece and a perturbed Kitaev model $H_K[h_\mathrm{mf}] \equiv H_K - \sum_i \vec{h}_\mathrm{mf} \cdot \vec{\sigma}_i$ with an effective mean Zeeman field given by $\vec{h}_\mathrm{mf} \equiv - J \vec{m}$, which is to be determined by demanding self-consistency.
A finite field $\vec{h} \neq 0$ spoils the exact solvability of the Kitaev model, but we may leverage recent developments in using controlled perturbative analysis to describe the dynamics of excitations~\cite{prxvison,inti,baskaran07,kitaevpowerlaw}.

In the perturbative regime, at $T=0$, we obtain the electronic magnetization $\vec{m} = - J  \vec{M} \chi_0 $ as a response to the local-moment magnetization $\vec{M}$ using the static susceptibility $\chi_0=2 |E_F|/(\pi v_\mathrm{F}^2)$ (here, $E_F$ is the Fermi energy, $v_\mathrm{F}\approx\sqrt{3}|t|a/2$ is the Fermi velocity on the honeycomb lattice near half-filling).
Eliminating $\vec{M}$, the energy functional reads $E_\mathrm{mf}[\vec{m}] = E_\mathrm{mf}[0] + \delta E_\text{e}[\vec{m}] + \delta E_\mathrm{K}[\vec{m}]$, with the electronic contribution $\delta E_\text{e}[\vec{m}] = N |\vec{m}|^2 / (2\chi_0)$.
To obtain the energy of the spin liquid $\delta E_\mathrm{K}[\vec{m}]$ perturbed by a mean field $\vec{h}_\mathrm{mf} = - J \vec{m}=-J(m^x,m^y,m^z)$, we note that a single operator $\sigma^\alpha_i$ creates two vison excitations on adjacent plaquettes sharing an $\langle ij\rangle_\alpha$ bond and extra Majorana fermions~\cite{baskaran07}.
Hence, the leading-order correction to the ground state energy corresponds to a second-order process where the same pair of visons is created and annihilated, yielding $\delta E_{\text{SL}}/N \approx -\eta J^2m^2/\Delta_{v}$ where $m=|\vec m|$.
The numerical factor $\eta\approx 0.434$ is obtained from the exact static spin susceptibility of the Kitaev model~\cite{knolle, bunneyPRL}.

Combining the electronic and spin liquid systems, the $\vec{m}$-dependent contribution to the mean-field energy functional (at leading order) is therefore given by
\begin{equation} \label{eq:kit-kondo-gs-mf}
    \delta E_\mathrm{mf}[\vec{m}]\approx {Nm^2}\left(\frac{1}{2\chi_0}-\eta \frac{J^2}{\Delta_v}\right).
\end{equation}
Thus, no generic weak-coupling instability occurs, and instead some critical coupling $J>J_{c,0}\approx \sqrt{\Delta_v/(2\eta\chi_0)}$ is required for a magnetic instability.

Let us now consider a non-equilibrium state where a small but finite density of visons is injected into the system.
At $T=0$, this could be done, in principle, by coherent optical or magnetic driving~\cite{jin23,joyRaman}, using local manipulation~\cite{liu22,Egger}, or by a quench protocol~\cite{kqslspintransport}.
A finite magnetization $\vec{m}$ in the metallic layer acts like a Zeeman field which breaks the flux conservation law. Thus, visons would no longer remain immobile and instead become dispersing quasiparticles.
We now show that it is generically preferable to develop a nonzero magnetization in order to endow the visons with the maximal kinetic energy.

For sufficiently low vison densities, we can work in a single-particle approximation and obtain the ensuing energy dispersion of the vison, moving in a background of gapless Majorana fermions, using perturbation theory in the effective Zeeman mean field $\sim J\sum_i\vec m\cdot\vec \sigma_i$.
This was studied in detail in Refs.~\onlinecite{prxvison} and \onlinecite{inti}: For a weak field $\vec m$, a vison hops across an $\langle ij\rangle_\alpha$-bond with a hopping amplitude $Jm^\alpha t_v$ where $t_v=\bra{\Phi_0(\bvec{r}_v)}(\sigma^\alpha_i+\sigma^\alpha_j)\ket{\Phi_0(\bvec{r}+\bvec{\delta}_i^\alpha)}$, where $\ket{\Phi_0(\bvec{r}_v)}$ is the many-body ground state of the Majorana fermions with a vison at position $\bvec r_v$.
For a ferromagnetic Kitaev model, $t_v \approx -0.6 ~\text{sgn}(\kappa)$, where $\kappa \sim (m^xm^ym^z)J^3/K^2$ is the topological Majorana gap arising from time-reversal breaking.
Thus, the vison energy can be approximately described by a tight-binding dispersion $E_v(\bvec k,\vec m) \approx \Delta_v-J \epsilon_\bvec{k}$, where $\epsilon_\bvec{k}(\vec m) =-2t_v\sum_{\alpha} m^\alpha\cos{\bvec{k} \cdot \bvec \delta^\alpha}$. The vectors $\bvec{\delta}^\alpha, \alpha=x,y,z$ denote the three nearest neighbor bonds of the triangular lattice formed by the hexagonal plaquettes. 

Assuming that a small density $n_v$ of visons occupy the minimum of the dispersion,
the mean-field energy of the Kitaev-Kondo system can be written as
\begin{align}
\label{eq:kk_total_energy}
    E[m]\approx &Nm^2\left(\frac{1}{2\chi_0}-\frac{\eta J^2}{\Delta_v}\right)\\ \nonumber
    &+Nn_v\left(\min\{E_v(\bvec k,\vec m)\}-\mathcal{O}\left(J^2m^2/\Delta_v\right)\right). 
\end{align}
The vison band minimum can be obtained as $\min\{E_v(\bvec k,\vec m)\}=\Delta_v-2|t_v|\sum_\alpha |Jm^\alpha|$.
The energy functional is minimized by eight symmetry-equivalent configurations $\vec m=m_0(\pm1,\pm1,\pm1)$, which are related by discrete joint lattice- and spin rotations, and time reversal. This yields minima of the dispersion at $\Gamma$ and M points in the Brillouin zone. The spontaneous selection of one of these states leads to a finite magnetization of the itinerant electrons.
Thus, when a small but finite density of visons is present, their parametrically large gain in kinetic energy, linear in $|J m^\alpha|$, outweighs the quadratic energy cost for the electrons to spontaneously magnetize.

\begin{figure}
    \centering
    \includegraphics[width=\columnwidth]{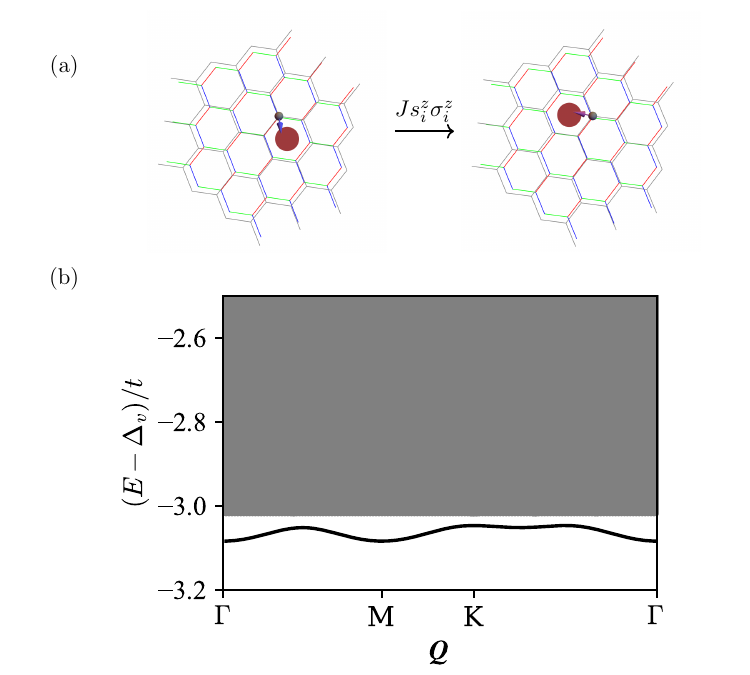}
    \caption{(a) The Kondo interaction $J \vec{s}_i \cdot \vec{\sigma}_i$ hops a vison, e.g., across a $z$-type bond (red) while simultaneously acting on the electron spin (arrow) with $s^z$ operator. (b) The energy spectrum obtained by solving the effective vison-electron pair problem (see Supplemental Material~\cite{suppl} for details), as a function of the center-of-mass momentum $\bvec Q$. A bound state emerges below and above (not shown) the continuum (gray shaded region). Parameters used: $t=1, g=Jt_v=0.2$.}
    \label{fig:vison_electron}
\end{figure}
{\em Beyond the mean-field approximation: analysis of RKKY interactions.---}%
Note that also in the absence of spontaneous symmetry breaking, the Kondo coupling $H_J$ breaks the $\mathbb{Z}_2$ 1-form symmetry and leads to vison dynamics.
We now argue that these effects are parametrically small compared to the kinetically-driven symmetry breaking discussed above.
To this end, we integrate out the itinerant electrons, which at lowest (quadratic) order yields a long-ranged Ruderman-Kittel-Kasuya-Yosida (RKKY) interaction between the spins of the Kitaev model~\cite{rkky2d}
\begin{align}
    H_{\text{RKKY}}= \sum_{i<j}-J^2\chi_0(\bvec r_i-\bvec r_j)\vec \sigma_i\cdot \vec \sigma_j,
    \label{eq:rkky}
\end{align}
where we have neglected retardation effects.
Then, the effect of $H_{\text{RKKY}}$ on single-vison states of $H_K$ may be studied perturbatively.
As detailed in the Supplemental Material~\cite{suppl}, using insights on the mobility of visons due to Heisenberg interactions~\cite{prxvison,lowenergyspinbalents}, we argue that single-vison motion due to $H_{\text{RKKY}}$ is strongly suppressed due to interference effects at least up to order $\mathcal O\left(J^4E_F^2/(K^2t^4)\right)$.
Thus the RKKY interaction is expected to only weakly affect our mean-field results~\cite{suppl}.
An additional subtlety arises from the power-law decay of the RKKY interaction which implies that a bound-pair of visons on two neighboring plaquettes acquires a long-ranged hopping amplitude~\cite{Batista}, but we argue that these excitations remain gapped for sufficiently small $J$~\cite{suppl}.

{\em Vison-electron bound state in the single-electron limit.---}%
We complement our analysis by investigating the dilute limit of a single conduction electron.
Within the single-vison sector, the Kondo coupling induces vison motion through spin-flip-assisted hopping processes linear in $J$, see Fig.~\ref{fig:vison_electron}(a). We diagonalize the single \mbox{(dressed-)}vison-electron Hamiltonian to obtain bound states below (and above) the continuum, as shown in Fig.~\ref{fig:vison_electron}(b).
The binding energy follows $E_b \sim e^{-ct/|J|}$ as expected for a weak local interaction in $2d$ where $c$ is a constant, see also Ref.~\onlinecite{suppl} for details.
The minima of the bound-state dispersion occur at the $\Gamma$ and M-points of the Brillouin zone.
The ground state is thus a generic superposition of the degenerate minima, entangling the spin and momentum of the vison-electron bound state, as required by joint spin-lattice rotational symmetries of the Kitaev system. 


{\em Order from thermal activation.---}%
While one could devise non-thermal protocols to induce a small density of excitations, a natural way to introduce them is by thermal activation. In contrast to the $1d$ model discussed previously, a spontaneous breaking of discrete symmetries at $T>0$ is allowed in the Kitaev-Kondo model.

In the dilute density limit, one can derive an effective free-energy functional for the system, which for $|Jm|< T< \Delta_v$ can be expanded in powers of $\eu^{-\beta \Delta_v}\ll 1$.
The magnetic contribution due to the itinerant fermions is obtained from the finite-$T$ spin-susceptibility $\chi(T)\sim \chi_0\left(1+\mathcal{O}(T^2/E_F^2)\right)$, and for sufficiently low temperatures we may neglect the temperature dependence.

The free energy of the spin liquid in the presence of a finite magnetization receives contributions from both gapless Majorana fermions and gapped vison excitations.
If the latter are dilute, the linear spectrum of Majorana fermions is only weakly affected~\cite{perkins}, and we may thus approximate the spin liquid's free energy by the sum of the two contributions, $F_{\text{SL}}\approx F_v+F_m$, where $F_v$ and $F_m$ are the free energies of the visons and Majoranas, respectively.
The leading temperature dependence of $F_m$ arises at $O(T^3/K^2)$ and can therefore be neglected for $T\ll K$, and we compute $F_v$ using the perturbative vison dispersion relation obtained earlier.

For a low-temperature expansion of $F_{\text{SL}}=-\beta^{-1}\ln Z_{\text{SL}}$, we use that the partition function $Z_{\text{SL}}$ at temperature $T=1/\beta$ can be written as
\begin{equation}
    Z_{\text{SL}} \approx \eu^{\beta N \frac{\eta J^2m^2}{\Delta_v}}+\sum_{ \bvec{k} \in\text{BZ}} \eu^{-\beta\left(\Delta_v-J \epsilon_{\bvec k}-T\ln\sqrt{2}+\mathcal{O}\left(\frac{ J^2m^2}{\Delta_v}\right)\right)}.
\end{equation}
The $\mathcal{O}(J^2/\Delta_v)$ terms in the exponential  encode the second-order corrections to the ground and excited states, arising from pair creation-annihilation of visons and Majorana fermions.
The extra entropic factor $T\ln\sqrt 2$ arises from the Majorana zero modes attached to visons in the time-reversal broken phase (see also Appendix F of Ref.~\onlinecite{prxvison}).

Up to quadratic order in $J$, we obtain
\begin{align}
\label{eq:fe_spinliquid1}
    \frac{F}{N} \approx \frac{m^2}{2\chi_0}&-\frac{\eta J^2m^2}{\Delta_v}\\ \nonumber
    &-\sqrt{2} \eu^{-\beta\Delta_v}\left[\beta^{-1}+\frac{\beta J^2m^2}{2}A+\mathcal{O}\left(\frac{ J^2m^2}{\Delta_v}\right)\right].
\end{align}
where $A=t_v^2 V_{\mathrm{u.c.}} (2\pi)^{-2} \int_\text{1. BZ} \du^2 k \, \cos^2(\bvec k\cdot\bvec \delta^\alpha)$.
The itinerant fermions thus develop an instability toward finite magnetization when the coefficient of the harmonic term $m^2$ changes sign.
This occurs upon increasing $J$ and, remarkably, also by increasing $T$.

From Eq.~\eqref{eq:fe_spinliquid1}, we obtain the general result Eq.~\eqref{eq:relative-gc-T}, which also applies to the dipole chain in the mean-field approximation.
This suggests that the critical coupling $g_c$ {\em decreases} as $T$ increases, with a lower bound $g_{c,0}/\sqrt{1+A/\eu}$, as shown in Fig.~\ref{fig:introfig}(c).
We interpret this as a kinetically-driven order-by-disorder mechanism wherein the degeneracy of local mobility-constrained excitations is lifted by undergoing symmetry breaking.

We caution that the minimal value of $g_c$ is realized when $T\approx \Delta$. At this point, the dilute-vison approximation breaks down, and vison-vison interactions will become important.
Nevertheless, we argue that the reduction of the critical coupling with increasing $T$ is a robust prediction for temperatures much smaller than $\Delta$.


{\em Discussion.---}%
We have described a mechanism for long-range ordering, driven by the gain in kinetic energy of mobility-constrained excitations due to condensation of a local order parameter.
Beyond the limit of strictly immobile quasiparticles,~e.g., when weak non-Kitaev terms provide the visons with a bandwidth $W_0$, we expect the kinetic mechanism to become important when $W_0$ is the smallest energy scale in the system, $W_0\ll J\ll T$.
Furthermore, the possible interplay or competition of the outlined mechanism with superconducting and Kondo instabilities~\cite{bunneyPRL,seifertvojta,choiRosch} is left for future work.
It would further be interesting to explore whether suppressed quasiparticle motion due to emergent dynamical constraints can stabilize unconventional order, including topological order~\cite{castelnovorvb}.

Experimentally, the mechanism could be explored in programmable quantum simulators on which,~e.g.,~the Kitaev model has been realized~\cite{evered2025probing}.
Here, non-equilibrium state preparation protocols will produce a finite density of visons.
A complementary route is offered by dipolar-Hubbard and cavity platforms~\cite{kim2025multi}, where cavity-mediated correlated hopping can relax dipole constraints, and the kinetic mechanism discussed here could induce symmetry-breaking of the cavity field itself~\cite{rylands2020photon}.

{\em Acknowledgements.---}%
We thank Sebastian Granberg Chauchi, Achim Rosch and Matthias Vojta for insightful discussions and for critical feedback on the manuscript.
This work is funded by the Deutsche Forschungsgemeinschaft (DFG, German Research Foundation) through SFB 1238, project ID 277146847 (APJ and UFPS), and the Emmy Noether Program, project ID 544397233, SE 3196/2-1 (UFPS).

\bibliography{references}

\end{document}


\title{Supplemental Material to ``Kinetically induced order from mobility-constrained excitations''}

\author{Aprem P. Joy}
\author{Urban F.P. Seifert}
\affiliation{%
	Institute for Theoretical Physics, University of Cologne, Cologne, Germany
}%
\date{\today}

\maketitle

\tableofcontents

\section{Dipole spin chain}

\subsection{Low-lying eigenstates}

We consider $H = H_D + H_J$ as given in Eq.~(1) of the main text.

\subsubsection{Ground state}

For $D \gg J > 0$, the system has a unique ground state $\ket{0} = \prod_i \ket{0_i}$. We stress that $\ket{0}$ is an exact eigenstate at any $J/D$, with energy $E_0 =0$.

\subsubsection{Single-particle states}

The lowest-lying set of excited states is given by \emph{single-particle states}.
To lowest order in $J/D$, they are obtained by acting with a spin-raising/lowering operator on the vacuum, $\ket{(\pm,i)} = \ket{ \dots 0_{i-1} \pm_i 0_{i+1} \dots } + \mathcal{O}(J/D)$.
The energy of these states is obtained as $E_\mathsf{s} = +D + \mathcal{O}(J^2/D)$. 

Note that perturbative corrections arise from coupling to the three-particle sector, since
\begin{equation} \label{eq:single-particle-hj}
    H_J \ket{\dots 0_{i-1} \pm_i 0_{i+1} \dots 0} = - 4 J \ket{\dots \pm_{i-1} \mp_i \pm_{i+1} \dots }.
\end{equation}
We stress that these perturbative corrections \emph{do not} allow for mobility of excitations, which remains constrained by the symmetry of conservation of magnetization dipoles.

With Eq.~\eqref{eq:single-particle-hj} and the bare energy difference $+2D$ between one- and three-particle sectors, we can give the dressed single-particle states (to first order in $J/D \ll 1$) as
\begin{equation} \label{eq:single-particle-dressed}
    \ket{(\pm,i)^{(1)}} = \ket{\dots 0_{i-1} \pm_{i} 0_{i+1}} + \frac{2J}{D} \ket{\dots 0_{i-2} \pm_{i-1} \mp_{i} \pm_{i+1} 0_{i+2}} + \mathcal{O}\big((J/D)^2\big)
\end{equation}
and the excitation energy to second order as
\begin{equation} \label{eq:e-s}
    E_{\mathsf{s}} = + D - \frac{8J^2}{D} + \mathcal{O}\big(J^3/D^2\big)
\end{equation}

\subsubsection{Two-particle excitations}

Turning to two-particle states, we note that many of them can be thought of as product states of two immobile single-particle excitations with energy $+2D + \mathcal{O}(J^2/D)$.

A notable exception is given by \emph{dipole states} as introduced in the main text, which we label as $\ket{(\leftarrow,i)} = \ket{\dots 0_{i-1} +_i -_{i+1} 0_{i+2}}$ and $\ket{(\rightarrow,i)} = \ket{\dots 0_{i-1} -_i +_{i+1} 0_{i+2}}$.
Here, the arrow indicates the positive/negative dipole charge, $P^z \ket{(\leftarrow,i)} = (-1)  \ket{(\leftarrow,i)}$ and $P^z \ket{(\rightarrow,i)} = (+1)  \ket{(\rightarrow,i)}$.
Importantly, upon acting with $H_J$ on one of these dipole states, one remains within the sector of dipole states,
\begin{equation}
    H_J \ket{(\leftarrow,i)} = -4 J \ket{(\leftarrow,i+1)} -4 J \ket{(\leftarrow,i-1)},
\end{equation}
and similarly for $\ket{(\rightarrow,i)}$.
We may thus construct \emph{exact eigenstates} of $H_J + H_D$ as
\begin{equation} \label{eq:d-k}
	\ket{(\mathsf{d},k)} = \frac{1}{\sqrt{N}} \sum_{j=0}^{N-1} \eu^{\iu k j} \ket{(\mathsf{d},j)}
\end{equation}
where $\mathsf{d}= \leftarrow,\rightarrow$ and, for periodic boundary conditions, the momenta are given by $k =2\pi m /N$ with $-N/2 < m \leq N/2$.
The corresponding dispersion reads $\varepsilon_\leftarrow(k) = \varepsilon_\rightarrow(k) = - 8 J \cos k$.

\subsection{Perturbative mean-field analysis at zero temperature $T=0$}

We consider the expectation value of $E_\text{el} + H_g[u] + H$ to derive an energy functional $E[u]$ for $u$.
Assuming that $\ket{\psi[u]}$ is the ground state of the Hamiltonian $H_\text{g}[u] + H$, we compute 
\begin{equation}
	\braket{ \psi[u] | H_g[u] + H | \psi[u] } = E^{(0)} + E^{(1)} + E^{(2)} + \dots
\end{equation}
perturbatively in $g u \ll D,J $.

\subsubsection{Ground state}

Consider first that the system is in the ground state $\ket{0} \equiv \ket{0\dots 0}$ for $g=0$.
Because $\braket{0 | \vec{S}_i \cdot \vec{S}_{i+1} | 0} = 0$, the first-order correction vanishes.
For the second-order contributions, consider the action of $H_g[u]$ on $\ket{0}$, which yields
\begin{equation} \label{eq:s-ph-on-0}
	H_g[u] \ket{0} = g u \sum_i (-1)^i \left( \ket{\dots 0_{i-1} +_{i} -_{i+1} 0_{i+2} \dots  } + \ket{\dots 0_{i-1} -_{i} +_{i+1} 0_{i+2} \dots  } \right) \equiv g u \sum_i \sum_{d=\leftarrow,\rightarrow} (-1)^i \ket{(\mathsf{d},i)}.
\end{equation}
Hence, there are only matrix elements from the ground state into the dispersive manifold of single-dipole states.
The corresponding contribution at second-order perturbation theory reads
\begin{equation} \label{eq:e0-2}
	E^{(2)}_0 
	= - \sum_{-\pi < k \leq \pi} \sum_{\mathsf{d}' = \leftarrow,\rightarrow} \frac{1}{2D + \varepsilon_{d'}(k)}  | \braket{(\mathsf{d}',k)| H_g[u] |0\dots 0}|^2 = - 2 N \times \frac{(g u)^2}{2 D + 8 J} = - N \times \frac{(gu)^2}{D} \left( 
    1-\frac{4 J}{D} + \mathcal{O} \big( J^2/D^2\big) \right).
\end{equation}
Here, we have used that with Eq.~\eqref{eq:s-ph-on-0}, the matrix element is evaluated as
\begin{equation}
	|\braket{(d',k)| H_g[u] |0}|^2 = (g u)^2 \bigg| \sum_{\mathsf{d}=\leftarrow,\rightarrow} \bra{(\mathsf{d}',k)} \sum_i (-1)^i \ket{(\mathsf{d},i)} \bigg|^2 = N (g u)^2 \delta_{k,\pi},
\end{equation}
and the dispersion of single-dipole states in Eq.~\eqref{eq:d-k}, which, evaluated at momentum $k= \pi$, yields $\varepsilon_{d'}(k=\pi) = 8J$.

We thus find for the energy functional 
\begin{equation} \label{eq:gs-e-final}
	E[u] = N \left(\frac{\rho}{2} - \frac{g^2}{D + 4J} \right) u^2 \approx N \left(\frac{\rho}{2} - \frac{g^2}{D} \right) u^2
\end{equation}
and there is no generic weak-coupling instability (for $g \ll J \ll D$): the gap protects the system. As a result, within the mean-field approximation, the dipole conservation symmetry remains unbroken up to some finite coupling strength.

\subsubsection{Excited sector}

We proceed as above and derive an energy functional in the $2N$-fold degenerate manifold of single-particle states perturbatively in $gu \ll D,J$.
To this end, consider the action of $H_g[u]$ on the dressed single-particle state in Eq.~\eqref{eq:single-particle-dressed}:
\begin{multline}
	H_g[u] \ket{(\pm,j)} = g u \left[ (-1)^{j-1} \ket{(\pm,j-1)} + (-1)^j \ket{(\pm,j+1)} + \sum_{i \neq j,j-1} (-1)^i \Big( \ket{(\pm,j);(\leftarrow,i)} + \ket{(\pm,j);(\rightarrow,i)} \Big) \right] \\
    + g u \frac{2J}{D} \Bigg[ (-1)^j \ket{(\pm,j-1)} + (-1)^{j-1} \ket{(\pm,j+1)} + (-1)^j\ket{\dots 0_{j-3} \pm_{j-2} 0_{j-1} \mp_j \pm_{j+1} 0_{j+2} \dots} \\
    + (-1)^{j+1} \ket{\dots 0_{j-2} \pm_{j-1} \mp_{j} 0_{j+1} \pm_{j+2} 0_{j+3} \dots } + \left(\text{5-particle states} \right) \Bigg]
\end{multline}
where $\ket{(+,j);(\leftarrow,i)} = \ket{\dots 0_{j-1} +_j 0_{j+1} \dots 0_{i-1} +_i -_{i+1} 0_{i+2} \dots }$ denotes a state with a localized single-particle excitation and a (mobile) dipole excitation.
Thus, $H_g[u]$ has finite matrix elements \emph{within} the (dressed) single-particle manifold, leading to a finite first-order contribution which we obtain by diagonalizing
\begin{equation} \label{eq:h-sph-subspace}
	\braket{(\mathsf{s},i)| H_g[u]| (\mathsf{s}',j)} = gu \left(1-\frac{2J}{D} \right) \, \delta_{\mathsf{s},\mathsf{s}'} \left( (-1)^{j-1} \delta_{i,j-1} + (-1)^j \delta_{i,j+1} \right).
\end{equation}
Henceforth, we introduce the renormalized coupling $\tilde{g} = g(1-2J/D)$, where higher-order contributions in $J/D \ll 1$ would arise from higher-order perturbative corrections to the single-particle states in Eq.~\eqref{eq:single-particle-dressed}.

Note that Eq.~\eqref{eq:h-sph-subspace} is translation invariant and we can thus introduce momentum states
\begin{equation} \label{eq:s-single-k}
	\ket{(\mathsf{s},k)} = \frac{1}{\sqrt{N}} \sum_j \eu^{\iu k j} \ket{(\mathsf{s},j)}
\end{equation}
with momentum $k$, which due to the staggered nature of interactions is only conserved modulo $\pi$, i.e. $\ket{(\mathsf{s},k)}$ and $\ket{(\mathsf{s},k+\pi)}$ hybridize.
Then the effective Hamiltonian at momentum $k$ is given by $\underline{\underline{H}}(k) = [2 (\tilde{g} u) \sin k ] \tau^y$, from which the single-particle states are seen to acquire a dispersion with two bands
\begin{equation} \label{eq:s-disp}
	\varepsilon_1(k) = - 2 |\tilde{g} u \sin k| \quad \text{and} \quad \varepsilon_2(k) = +2 | \tilde{g} u \sin k|,
\end{equation}
where $k$ is now restricted to the reduced Brillouin zone $ - \pi/2 < k \leq \pi/2$. Note that Eq.~\eqref{eq:h-sph-subspace} does not lift the degeneracy between $\mathsf{s} = \pm$, such that each state has a twofold degeneracy.
The dispersion in Eq.~\eqref{eq:s-disp} is gauge-equivalent (momentum shift by $\pi/2$) to zone-folding the dispersion
\begin{equation}
    \varepsilon(\tilde{k}) = - 2 |\tilde{g}| |u| \cos \tilde{k}
\end{equation}
with $-\pi < \tilde{k} \leq \pi$.
Note that higher-order corrections will arise at order $g^2/D$ from the coupling between the momentum eigenstates Eq.~\eqref{eq:s-single-k} and three- and five-particle states.

From Eq.~\eqref{eq:s-disp}, the lowest-lying one-particle state has energy $E_\mathsf{s}+\varepsilon_1(k=\pi/2) = -2 |\tilde{g} u| +E_\mathsf{s}$ with $E_\mathsf{s}$ given in Eq.~\eqref{eq:e-s} which is an overall additive constant in the one-particle sector.
This yields the first-order correction to the system's energy in the one-particle sector,
\begin{equation} \label{eq:e-s-1}
	E_{\mathsf{s}}^{(1)}[u] = - 2 |\tilde{g} u|. 
\end{equation}

\subsection{Perturbative mean-field analysis at finite temperatures $T >0$}

Working in a mean-field approximation for the field $u$, the system's partition function at temperature $T = \beta^{-1}$ is given by
\begin{equation}
    \mathcal{Z}(\beta) = \int \mathcal{D}[u] \eu^{-\beta E_\mathrm{el.}[u] - \beta F_\mathrm{mag.}[u]}
\end{equation}
where $F_\mathrm{mag.}[u] = - \beta^{-1} \ln Z_\mathrm{mag.}$ is the free energy of the magnetic sector, which can be obtained from the partition function $Z_\mathrm{mag.} = \tr[\eu^{-\beta (H_D + H_J + H_g[u])}]$.
At low temperatures $T\ll D$, we may perform a low-temperature expansion of $F_\mathrm{mag.}$ in $\eu^{-\beta D} \ll 1$. For this, it is useful to expand the partition function  to linear order
\begin{equation}
    Z_\mathrm{mag.} = \eu^{-\beta E_0} + 2 \sum_{-\pi < \tilde{k} \leq \pi} \eu^{-\beta (E_\mathrm{s} + \varepsilon(\tilde{k}))} + \mathcal{O}\left(\eu^{-2\beta  D}\right).
\end{equation}
In the regime $g \ll J \ll D$, we use the perturbatively obtained eigenenergies in Eq.~\eqref{eq:e0-2} and Eq.~\eqref{eq:e-s-1}, yielding
\begin{equation}
    Z_\mathrm{mag.} = \eu^{\beta N (gu)^2/D} + 2 \sum_{-\pi <\tilde{k} \leq \pi} \eu^{-\beta (D - 8J^2/D - 2 |\tilde{g} u| \cos \tilde{k} + \mathcal{O}(|gu|^2/D))} + \mathcal{O}(\eu^{-2 \beta D}).
\end{equation}
We thus obtain the free energy as
\begin{align}
    - \beta F_\mathrm{mag.}[u] = \ln Z_\mathrm{mag.} &= \beta N \frac{(gu)^2}{D} + \ln\left( 1 +  2 \eu^{-\beta (D-8J^2/D)} \eu^{-\beta N (gu)^2/D} \sum_{-\pi<\tilde{k}\leq \pi} \eu^{2 \beta |\tilde{g} u| \cos \tilde{k} + \mathcal{O}(\beta |gu|^2/D)} + \mathcal{O}(\eu^{-2\beta D}) \right) \nonumber\\
    &= \beta N \frac{(gu)^2}{D} + 2 \eu^{-\beta D} \left(\eu^{-\beta (N (gu)^2-8J^2)/D} \sum_{-\pi<\tilde{k}\leq \pi} \eu^{2 \beta |\tilde{g} u| \cos \tilde{k} + \mathcal{O}(\beta |gu|^2/D)}  \right) + \mathcal{O}(\eu^{-2\beta D}) \label{eq:f-k-sum}
\end{align}
The momentum sum in Eq.~\eqref{eq:f-k-sum} may be evaluated in the continuum limit upon writing $2 \sum_k \eu^{2\beta |g| |u| \cos k} \approx 2N (2\pi)^{-1}  \int_0^{2\pi} \du k \,  \eu^{2\beta |g| |u| \cos k} = 2N I_0(2 \beta |g| |u|)$, where $I_0(x)$ is the $0$-th modified Bessel function of the first kind.
More insight is obtained by expanding to second order in $\beta |g u| \ll 1$ (and first order in $\beta |gu|^2 /D \ll 1$ which is consistent provided that $D\gg |g u|$), noting that $\int_{-\pi}^\pi  \du \tilde{k} \cos  \tilde{k}  = 0$,
\begin{equation}
    - \beta F_\mathrm{mag.}[u] \approx \beta N \frac{(gu)^2}{D} + 2 \eu^{-\beta D} \eu^{8 \beta J^2/D} \left[ \left(1 -\beta N\frac{(gu)^2}{D} \right) \times \left(N + N\mathcal{O}(\beta |gu|^2/D) + \frac{1}{2} (2\beta |gu|)^2 N \int_{-\pi}^\pi \frac{\du \tilde{k}}{2\pi} \cos^2 \tilde{k} \right)  \right].
\end{equation}
We thus arrive at the free energy density (we take $\eu^{8 \beta J^2/D} \approx 1$ for simplicity)
\begin{equation}
    N^{-1} F_\mathrm{mag.}[u] \approx \left[-\frac{g^2}{D} \left(1 + \eu^{-\beta D} \mathcal{O}(1) \right) - 2 \beta g^2 \eu^{-\beta D}\right] u^2 - \frac{2}{\beta} \eu^{-\beta D},
\end{equation}
where $\mathcal{O}(1)$ denotes a numerical constant of arbitrary sign.
An instability occurs once the quadratic coefficient of  $E_\mathrm{el.}[u] + F_\mathrm{mag.}[u]$ with $E_\mathrm{el.}[u] = N \rho u^2/2$ changes sign, from which we determine the critical coupling at finite $T= \beta^{-1}$ as
\begin{equation}
    \frac{\rho}{2} = \frac{g^2(\beta)}{D} \left[1 + \mathcal{O}(1) \eu^{-\beta D} + 2 \beta D  \eu^{-\beta D} \right] \quad  \Leftrightarrow \quad \frac{g^2(\beta)}{D} = \frac{\rho}{2} \frac{1}{1 + \mathcal{O}(1) \eu^{-\beta D} + 2 \beta D  \eu^{-\beta D}}
\end{equation}
We now expand in $\eu^{-\beta D} \ll 1$. Noting that $(\beta D) \eu^{-\beta D}$ is parametrically large compared to $\mathcal{O}(1) \eu^{-\beta D}$ for $\beta D \gg 1$, we thus arrive at
\begin{equation}
    \frac{g_c^2(T)}{D} \approx \frac{\rho}{2} \left(1 - 2 (D/T) \eu^{-D/T} \right) \quad \Leftrightarrow \quad g_c^2(T) - g_c^2(0) \approx - A g_c^2(0)  (D/T) \eu^{-(D/T)},
\end{equation}
where $g_c^2(0) = \rho D/2$ is the critical spin-lattice coupling for the system in the $T=0$ ground state, as obtained from Eq.~\eqref{eq:gs-e-final}, and $A =2$ is a numerical constant.

\section{Kitaev model and visons}

We briefly review the Kitaev honeycomb model and its emergent excitations. The Kitaev model is defined on a honeycomb lattice of spin-1/2 moments $\vec{S}_i = \vec\sigma_i/2$, with the Hamiltonian~\cite{Kitaev06}
\begin{equation}
    H_{K}=- K \sum_{\langle ij\rangle_\alpha} \sigma_i^\alpha \sigma_j^\alpha
\end{equation}
Thus, every site connects to its three nearest neighbors through an $\alpha=x,y,z$-type Ising interaction.
The model has an extensive number of conserved quantities called plaquette operators $W_p$, defined by a product of the Kitaev terms $\sigma_i^\alpha \sigma_j^\alpha$ around an elementary hexagon.
By introducing Majorana fermions of four types $c_i, b^x_i,b^y_i,b^z_i$, one can express every spin operator as $\sigma^\alpha_i = ic_ib^\alpha_i$.
This doubles the Hilbert space and transforms the spin Hamiltonian into a model of Majorana fermions hopping in the background of a $\mathbb Z_2$ gauge field. The gauge field variables are described by the variable $u^\alpha_{ij}=ib^\alpha_ib^\alpha_j$ for given type of bond.
The fermionic Hamiltonian thus reads
\begin{equation} \label{eq:h-k-fractionalized}
    H_K = K\sum_{\langle ij \rangle}i u^\alpha_{ij}c_ic_j.
\end{equation}
The exact eigenstates of the model can in principle be obtained by diagonalizing the quadratic Majorana fermionic Hamiltonian for a given configuration of the $\mathbb Z_2$ gauge variables $u_{ij}$. However, note that different choices of these variables may differ only by a choice of gauge, i.e., are physically equivalent.
Crucially, one may use the conserved plaquette operators, which can be rewritten in terms of the Wilson loop operators of the $\mathbb{Z}_2$ gauge field around a plaquette, $W_p = \prod_{\langle ij \rangle \in \Hhex} u_{ij}^\alpha$, to decompose the Hilbert space into physically distinct sectors, labeled by the eigenvalues $w_p = \pm 1$ of $W_p$ on each plaquette.
The model's ground state lies in the zero-flux sector $W_p = +1 \ \forall p$. One convenient choice of gauge for this sector consists in choosing $u_{ij}^\alpha = + 1$ for $i \in A$, $j\in B$ sublattice sites. Diagonalizing the Majorana fermion problem in this gauge, one obtains a gapless Dirac spectrum for the Majorana fermions with a velocity $v_m=\sqrt{3} aK/2$, where $a$ is the lattice constant.

Flipping $w_p = +1 \to w_p = -1$ on plaquette $p$ defines a quasiparticle excitation, dubbed ``vison'', with an energy gap of $\Delta_v \approx 0.15 |K|$.
The main feature of the Kitaev model that plays a crucial role in the mechanism elaborated in the main text is the conservation law $[H,W_p]=0 \ \forall p$. This renders the visons completely localized excitations. Thus, considering a single-vison state, there is a macroscopic degeneracy associated with choosing the position of this excitation.

\subsection{Magnetization from minimizing the vison energy}

Here, we derive the low-energy spin-polarized manifold of the Kitaev-Kondo system when a finite density of visons is present.
Consider the mean-field energy functional when a small density of visons occupy the band minimum
\begin{align}
    E[\vec m,\bvec k]\approx N\left[\frac{m^2}{2\chi_0}-\frac{\eta J^2m^2}{\Delta_v}+n_v\min(\Delta_v-J\epsilon_{\bvec k})\right].
\end{align}
Since $\epsilon_{\bvec k}=-2t_v\sum_\alpha m^\alpha\cos(\bvec k\cdot \bvec \delta^\alpha)$, the minimum occurs at momenta $\bvec k_{\min}$ for which $\epsilon_{\bvec k_{\min}}=-2|t_v|(|m^x|+|m^y|+|m^z|)$, which is realized for $\bvec k_{\min}$ located at the $\Gamma$ or M points of the Brillouin zone.
We thus minimize the following energy functional with respect to $\vec m$
\begin{align}
    E[\vec m,\bvec k]\approx N\left[\frac{m^2}{2\chi_0}-\frac{\eta J^2m^2}{\Delta_v}-2Jn_v|t_v|(|m^x|+|m^y|+|m^z|)\right]. \label{eq:kk_total_energy}
\end{align}
This gives the condition $|m^x|=|m^y|=|m^z|=m_0$, where $m_0=2Jn_v |t_v|\left(\frac{1}{2\chi_0}-\frac{\eta J^2}{\Delta_v}\right)^{-1}$.
We obtain an 8-fold degenerate manifold of energy minima:
\begin{align}
    \bvec k_{\min}=&(0,0), \quad \vec m \propto \pm(1,1,1)\\ \nonumber
    \bvec k_{\min}=&\text{M}_1, \quad \vec m \propto \pm(-1,1,-1)\\ \nonumber
    \bvec k_{\min}=&\text {M}_2, \quad \vec m \propto \pm(1,-1,-1)\\ \nonumber
    \bvec k_{\min}=&\text{M}_3, \quad \vec m \propto \pm(-1,-1,1).
\end{align}
where $\text{M}_1$, $\text{M}_2$ and $\text{M}_3$ are the three inequivalent M-points of the hexagonal Brillouin zone. Spontaneous symmetry breaking then chooses one of the above 8 states, and the electrons obtain a finite spin polarization.

\subsection{Effects of RKKY interaction}

\begin{figure}
    \centering
    \includegraphics[width=0.35\linewidth]{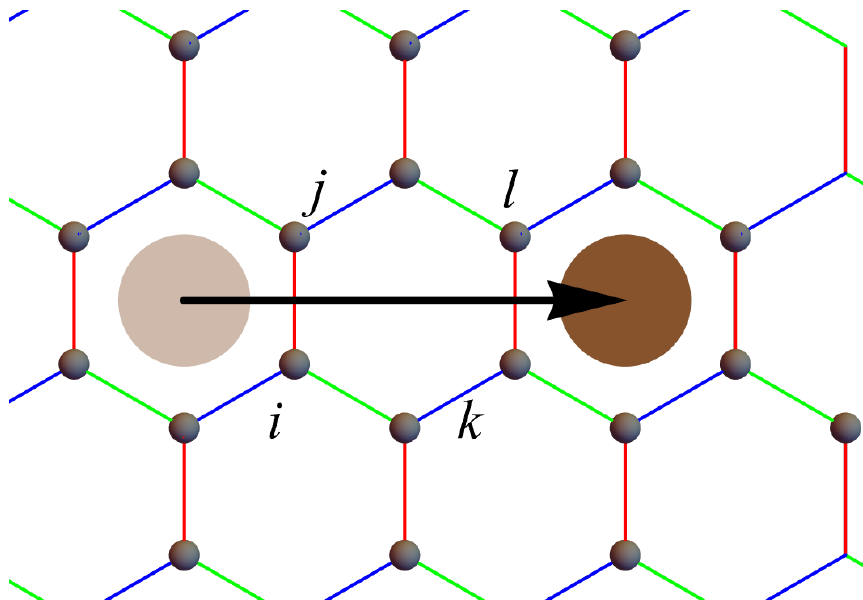}
    \caption{Vison hopping process induced by next-nearest- and third-nearest-neighbor terms of the RKKY interaction (Eq.~\eqref{eq:rkky}). Four different hopping processes $\hat T_1=\sigma^z_i\sigma^z_k$, $\hat T_2=\sigma^z_i\sigma^z_l$, $\hat T_3=\sigma^z_j\sigma^z_k$, $\hat T_4=\sigma^z_j\sigma^z_l$ contribute to the hopping (shown by arrow). The corresponding hopping matrix elements obtained by numerical calculation show $t_1=-t_4$ and $t_2=-t_3$.}
    \label{fig:rkky_hop}
\end{figure}
In this section, we discuss the effect of electron-mediated spin-spin-interactions on the Kitaev spin liquid, with a particular focus on dynamics of visons. Our goal is to investigate corrections to the disordered (trivial) mean-field regime $\vec{m} = 0$, i.e.~without any magnetization of the itinerant electrons.
It is well known that a Fermi liquid can generate an effective long-ranged Heisenberg coupling between local moments which are Kondo-coupled to the fermions via an effective RKKY interaction~\cite{rkky2d},
\begin{align}
    H_{\text{RKKY}}= -J^2\sum_{ij} \chi_0(\bvec r_i-\bvec r_j)\vec \sigma_i\cdot \vec \sigma_j,\label{eq:rkky}
\end{align}
where we approximate the interaction to be instantaneous (i.e.~neglect retardation effects), which is justified when the electronic degrees of freedom are sufficiently fast compared to the local-moment sector.

For the honeycomb lattice, this leads to an asymptotic form for large $r_{ij}=|\bvec r_i-\bvec r_j|$ (away from half-filling)
\begin{align} \label{eq:rkky-explicit}
H_{\text{RKKY}}\sim -\sum_{ij}J^2\rho(E_F)\frac{a^4\sin(2k_\mathrm{F}~r_{ij})}{r_{ij}^2}\vec \sigma_i\cdot \vec \sigma_j,
\end{align}
where $k_\mathrm{F}$ is the Fermi momentum of the itinerant electrons, $\rho(E_F)\sim 2|E_F|/(\pi v_F^2)$ is the density of states near the Fermi surface and $a$ is the lattice constant.

For the spin liquid, this interaction breaks the plaquette conservation law for any finite $J$. Nevertheless, below we show that single vison hopping due to weak Heisenberg interactions is strongly suppressed due to destructive interference of hopping processes.

\subsubsection{Single-vison dispersion}

We first focus on the nearest-neighbor part of the interaction, i.e., $i,j$ as neighboring sites.
At first-order perturbation theory in $H_{\text{RKKY}}$, this term does not induce any hopping of the vison since it maps a single vison state to a 3-vison or 5-vison state.
In fact, this is true for any odd-order process in perturbation theory.
At second order in $H_{\text{RKKY}}$, i.e., $\mathcal{O}(J^4)$, there are multiple processes that can hop a single vison to a third-nearest-neighbor plaquette, as shown in Fig.~\ref{fig:rkky_hop}. However, as was shown in Ref.~\cite{prxvison}, all such processes destructively interfere, resulting in a vanishing hopping amplitude.

Next, we consider the next-nearest-neighbor and third-nearest neighbor interaction terms. In this case, already at first order in $H_{\text{RKKY}}$, i.e., second order in $J$, one finds four terms that can hop a given vison to its next-nearest-neighbor plaquette. These are related to each other by an inversion about the center of the plaquette, as shown in Fig.~\ref{fig:rkky_hop}.
Using methods described in Ref.~\cite{prxvison}, we have computed the amplitude for all four processes using perturbation theory on a finite-size system. Importantly, we find that the four hopping amplitudes cancel each other pairwise, similar to the second-order processes discussed above.

Any further-neighbor interactions beyond next-nearest-neighbor cannot hop a single vison at linear order in $H_{\text{RKKY}}$, since (most of) such terms create extra vison pairs.
However, we remark that an interesting class of processes exists which involve $\vec{\sigma}_i$ and $\vec{\sigma}_j$ acting on sites that belong to two occupied vison plaquettes.
This leads to the simultaneous (correlated) hopping of two visons that are far away from each other. 
However, such processes -- symbolically written as $ n_v({\bvec r})n_v(\bvec r')v^\dagger_{\bvec r+\bvec \delta^\alpha}v_{\bvec r}v^\dagger_{\bvec r'+\bvec \delta^\alpha}v_{\bvec r'}$ -- are suppressed in the density of visons $n_v^2\ll 1$, for $T<\Delta_v$.

Thus we conclude that the RKKY interaction does not induce single vison motion at least up to $\mathcal{O}(J^3)$, and we therefore expect our mean-field results to enjoy an enhanced robustness in the Kitaev-Kondo model.

\subsubsection{Long-range hopping of vison pairs}

While above arguments show that $H_\mathrm{RKKY}$ only leads to a parametrically weak single-vison dispersion, the situation is more subtle when considering vison pairs on adjacent plaquettes.
As a single spin operator can annihilate/create such a pair, a Heisenberg interaction can lead to a finite hopping amplitude already in first-order perturbation theory.
Given the long-ranged nature of interactions in Eq.~\eqref{eq:rkky-explicit}, the vison-pair hopping can be long-ranged with a hopping amplitude $t_{\text{2v}}(\bvec r)\sim \frac{\sin(2k_Fr)}{r^2}$.
While this physics does not directly affect the single vison motion discussed above, long-range hopping of pairs can potentially destabilize the spin liquid by a proliferation of vison pairs.
We now argue that despite this long-ranged hopping, the vison pair remains gapped for sufficiently small $J$.

Consider a single vison-pair $\mathrm{vp}_\alpha$ with a common $\langle ij\rangle_\alpha$ bond at position $\bvec r=0$.
Since the RKKY interaction is SU(2)-symmetric, the three types of vison pairs ($\alpha=x,y,z$) do not hybridize~\cite{joy2024gauge}.
At first-order perturbation theory in $H_{\text{RKKY}}$, the term $\sigma_i^\alpha \sigma_k^\alpha$ annihilates $\mathrm{vp}_\alpha$ at $\bvec r=0$ and creates a pair at the bond $\langle kl\rangle_\alpha$, taken to be at position $\bvec r$.
The corresponding hopping amplitude is $t_{\mathrm{vp}_\alpha}(\bvec r)\sim \frac{J^2E_F}{t^2}\frac{a^2\sin(2k_\mathrm{F}r)}{r^2}$ with $r= |\bvec{r}|$.
The effective Hamiltonian of a single vison-pair can thus be described by a tight-binding model with long-range hopping $t_{\mathrm{vp}_\alpha}(\bvec r)$ on a square lattice formed by the $\alpha$-type bonds,
\begin{align}
    H_{\mathrm{vp}_\alpha} = \sum_{\bvec r_1,\bvec r}t_{\mathrm{vp}_\alpha}(\bvec r) \ket{\Phi_0(\bvec r_1+\bvec r)}\bra{\Phi_0(\bvec r_1)}+\Delta_{2v},
\end{align}
where $\Delta_{2v}\approx 0.27 |K|$ is the energy gap of a vison pair in the $J=0$ limit. Within our perturbative analysis, the vison pairs can proliferate when their band gap vanishes, signaling an instability of the spin liquid. Note that this does not directly lead to confinement.
\textit{A priori}, it is not obvious if this requires a finite hopping strength, given the long-ranged nature of hopping.

Using a continuum limit for $\bvec r$, we can estimate the dispersion and band gap of the vison pair by Fourier-transforming the hopping amplitude to $\bvec k$-space,
\begin{align}
    E_{\text{2v}}(\bvec k)=&\Delta_{2v}- \frac{J^2E_\mathrm{F} a^2}{t^2}\int  \mathrm{d}^2 \bvec r \frac{\sin(2k_\mathrm{F} r)}{r^2}\cos(\bvec k\cdot\bvec r)\\ \nonumber=&\Delta_{2v}-2\pi \frac{J^2 E_\mathrm{F} a^2}{t^2}\int_0^\infty J_0(kr)\frac{\sin(2k_\mathrm{F} r)}{r} \du r\\ \nonumber
    =&\begin{cases}
\Delta_{2v}-\frac{\pi^2J^2E_\mathrm{F} a^2}{t^2}, & k < k_\mathrm{F}, \\[6pt]
 \Delta_{2v}-\frac{2\pi J^2E_\mathrm{F} a^2}{t^2}\arcsin\!\left(\frac{k_\mathrm{F}}{k}\right), & k > k_\mathrm{F}.
\end{cases}
\end{align}
Thus, for small $J$, the vison pair remains gapped.

\subsection{Vison-electron Schrödinger equation}

Here, we present some details of the single vison-electron pair dynamics studied on a finite-size system.
The effective Hamiltonian of the vison-electron pair is described by the kinetic energy term of the spinful electron and the Kondo-coupling projected onto the single vison sector of the Kitaev spin liquid. For $J=0$, the vison is localized. At linear order in $J$, the Kondo interaction acts on a single vison by hopping it to a neighboring plaquette while flipping the spin of the electron along one of the axes.
For a single vison-electron pair, the effective interaction can be thus expressed as
\begin{equation}
    \hat V_{v-e}(\bvec r_v)=-Jt_v\sum_{\langle ij\rangle_\alpha}\ket{\Phi_0(\bvec r_v+\bvec \delta_\alpha)}\bra{\Phi_0(\bvec r_v)}(c^\dagger_{i,\rho}\tau^\alpha_{\rho\rho'}c_{i,\rho'}+c^\dagger_{j,\rho}\tau^\alpha_{\rho\rho'}c_{j,\rho'}).
\end{equation}
where the index $\langle ij \rangle_\alpha$ runs over the links that form the hexagon corresponding to the vison plaquette center $\bvec r_v$.
The total two-particle Hamiltonian is thus
\begin{align}
    H_{v-e}=-\sum_{s=\{\uparrow,\downarrow\},\langle ij\rangle}t(c^\dagger_{i,s}c_{j,s}+\text{h.c.})+\sum_{\bvec r_v} \left(\Delta_v\ket{\Phi_0(\bvec r_v)}\bra{\Phi_0(\bvec r_v)}+\hat V_{v-e}(\bvec r_v)\right)
\end{align}

Note that $\ket{\Phi_0(\bvec r)}$ describes a dressed-vison ground state where the Majorana fermion system is in a many-body ground state associated with a vison at position $\bvec r$.
The vison-electron interaction $V_{v-e}$ thus becomes a spin-flip assisted hopping of the dressed vison, whose matrix elements are obtained following the methods developed in Ref.~\onlinecite{prxvison}.
Here, a subtle complication arises from the gapless nature of Majoranas when considering first-order perturbation theory in $J$.
As pointed out in Refs.~\onlinecite{prxvison} and \onlinecite{inti}, the hopping matrix element of a vison due to a single Pauli operator suffers from large finite-size effects due to the gapless spectrum of Majoranas.
Therefore, we follow the same procedure as in Ref.~\onlinecite{inti}, where a small Majorana gap $\kappa$ is added while evaluating the hopping amplitudes, and subsequently take the limit $\kappa\to0$~\cite{nasu_ptep}. Note that such a mass gap can be introduced through a three-spin interaction term without spoiling the flux conservation laws~\cite{Kitaev06}.

A standard strategy to solve the above Hamiltonian is to transform to the center-of-mass and relative coordinate frames~\cite{joyRaman}. Since the total momentum is a good quantum number due to translation symmetry, we can solve the problem within the relative coordinate basis for a given total momentum $\bvec Q$.
Parameterizing the vison and electron positions, $\bvec r_v $ and $\bvec r_e$, respectively, using center-of-mass (COM) and relative coordinates: $\bvec R=(\bvec r_v+\bvec r_e)/2$ and $\bvec r=(\bvec r_v-\bvec r_e)$, we can use the following ansatz for the two-particle wavefunction:
\begin{align}
    \ket{\Psi_n(\bvec Q)} =\sum_\bvec Re^{i\bvec Q\cdot \bvec R}\phi_n(\bvec Q,\bvec r)\ket{\bvec r,\bvec R}
\end{align}
The wavefunction $\phi_n(\bvec Q,\bvec r)$ is obtained by diagonalizing the Hamiltonian after projecting it onto the relative coordinate basis. Within the relative coordinate basis, where we choose the vison position to be the origin, one obtains a tight-binding model on a honeycomb lattice with hopping amplitudes $t^{\text{AB}}_{rel}\sim te^{i\frac{\bvec Q\cdot \bvec{\delta}^{\text{AB}}}{2}}$ away from the origin, i.e., when the electron is far away from the vison.
On the six sites of the hexagonal plaquette around the origin, the interaction term $\hat{V}_{v-e}$ is activated and one obtains hopping between the same sublattice sites: $t^{\text{AA}}_{rel}\sim Jt_ve^{i\frac{\bvec Q\cdot \bvec{\delta}^{\text{AA}}}{2}}$ (similarly for BB hopping).

\begin{figure}
    \centering
    \includegraphics[width=0.5\linewidth]{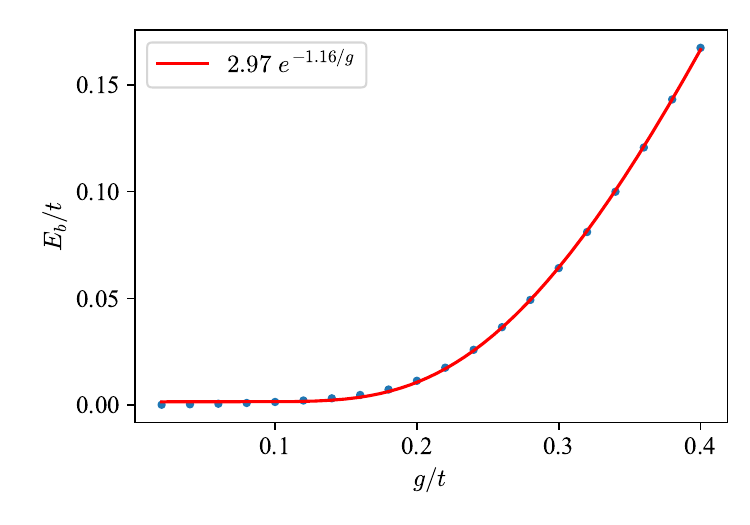}
    \caption{Vison-electron bound state energy extracted from exact diagonalization of the two-particle Schr\"odinger equation within the center-of-mass momentum $\bvec Q=0$ sector. The binding energy, as measured from the non-interacting band bottom, follows the exponential scaling expected for a weak interaction in $2d$, as shown by the fitted curve (red). Here, we have considered a finite-size cluster of $30\times30$ sites.}
    \label{fig:low_spectrum}
\end{figure}
For a given COM momentum $\bvec Q$, we can therefore obtain $\phi_n(\bvec Q,\bvec r)$ by diagonalizing this effective tight-binding model.
The vison-electron interaction thus generates an enhanced hopping around the central hexagon, and a kinetic bound state emerges.
As discussed in the main text, we obtain a dispersing bound state with its minima at the BZ center and at the midpoint of the BZ edges.
In Fig.~\ref{fig:low_spectrum}, we plot the binding energy of this state obtained for $\bvec Q=0$ as a function of the interaction strength $g$.
The lowest-energy state is two-fold degenerate and drops below the continuum edge as $g$ increases.
The scaling $E_b \sim e^{-t/g}$ implies a weakly (logarithmically) bound vison-electron composite, as expected for a weak local interaction in $2d$. 

\subsection{Free energy of a magnetized metal}

We consider itinerant fermions on a honeycomb lattice  with a chemical potential $\mu$. 
Upon applying a weak Zeeman field $\vec h$, the Fermi surfaces of the $\uparrow$- and $\downarrow$-spin electrons get shifted by $\pm h/2$.
The magnetic contribution to the free energy for $T\to 0$ can be written using the magnetic susceptibility $\chi(T,\mu)$ as
\begin{align}
    F_{\text{e}}=\frac{(m)^2}{2\chi(T,\mu)}.
\end{align}
At $T=0$, we can use $E_F=\mu$ where $E_F$ is the Fermi energy.
For electrons on a honeycomb lattice near (but \emph{not} at) half-filling, $\rho(E_F)=2|E_F|/(\pi v_F)^2$.

At $T>0$, the susceptibility acquires corrections which can be obtained to leading order in $T/E_F$ using the Sommerfeld expansion:
\begin{align}
    \chi_0(T)\approx \rho(E_F)\left[1-\frac{\pi^2}{6}\left(\frac{T}{E_F}\right)^2\right].
\end{align}

\subsection{Ground-state energy of Kitaev spin liquid perturbed by a weak Zeeman field}

We consider the spin liquid perturbed by an effective (mean) Zeeman field, given by the magnetization of the conduction electrons, $H_K+\sum_i J \vec{m} \cdot \vec{\sigma}_i$.
The first nontrivial correction to the ground state energy can be computed to second order in perturbation theory
\begin{align}
    \delta E_0 = -(Jm)^2\sum_{ij,\alpha}\sum_n\frac{\bra{0} \sigma^\alpha_i\ket{n}\bra{n}\sigma_j^\alpha\ket{0}}{E_n-E_0} (\delta_{ij}+\delta_{\langle ij\rangle_\alpha})
\end{align}
The onsite ($i=j$) and nearest-neighbor contributions were calculated to be approximately $0.67/|K|$ and $-0.59/|K|$ for $K>0$~\cite{bunneyPRL}, and we thus arrive at
\begin{align}
    \delta E_0 \approx -\eta\frac{N J^2m^2}{\Delta_v}
\end{align}
with $\eta\approx 0.434$.
The numerical factors account for the number of sites $N$ and bonds $3N/2$, and we use $\Delta_v\approx 0.15 |K|$.

\subsection{Free energy of Majorana fermions}

At finite temperature $T$, the gapless Majorana fermions also contribute to the free energy.
For $T\ll K$ one can approximate the Majorana spectrum by the Dirac dispersion $\epsilon_m(\bvec k)=v_m |\bvec{k}|$, where $v_m=\sqrt{3}|K|/2$, yielding
\begin{align}
    F_m=-T\int d k k\ln[1+e^{-\beta v_m k}]\propto -T^3/K^2
\end{align}
At second order in $h$, the renormalization of the Kitaev coupling modifies the Majorana velocity.
So for $h\ll K$, we can safely neglect the correction from the Zeeman term.

\bibliography{references.bib}

\bibliographystyle{apsrev4-2}